\documentclass[preprint, 11pt]{elsarticle}

\pdfoutput=1

\usepackage{natbib}
\usepackage{amsmath}
\usepackage{amstext}
\usepackage{amsfonts}
\usepackage{amssymb}
\usepackage{color}
\usepackage{graphicx}
\usepackage{hyperref}

\graphicspath{{graphics/}}

\newcommand{\Ref}[1]{Ref.~#1}
\newcommand{\Refs}[1]{Refs.~#1}
\newcommand{\Fig}[1]{Fig.~#1}
\newcommand{\Eq}[1]{Eq.~(#1)}

\title{\vspace*{-4cm}{\large \normalfont \hfill DESY 11-129, JLAB-THY-11-1389, SFB/CPP-11-44}\\\vspace*{3cm} 
Precision Study of Excited State Effects in Nucleon Matrix Elements}

\author[DZ]{Simon Dinter}
\ead{Simon.Dinter@desy.de}

\author[UC,TCI]{Constantia Alexandrou}

\author[UC]{Martha Constantinou} 

\author[DZ]{Vincent Drach}

\author[DZ]{Karl Jansen}
\ead{Karl.Jansen@desy.de}

\author[DZ,JL]{Dru B.\ Renner}

\author{\newline\\ for the ETM Collaboration}

\address[DZ] {NIC, DESY Zeuthen, Platanenallee 6, D-15738 Zeuthen, Germany}
\address[UC] {Department of Physics, University of Cyprus, P.O. Box 20537, 1678 Nicosia, Cyprus}
\address[TCI]{Computation-based Science and Technology Research Center,
              The Cyprus Institute,\\ 15 Kypranoros Str., 1645 Nicosia, Cyprus}
\address[JL] {Current address: Jefferson Lab.}

\biboptions{sort&compress}
\begin{document}

\begin{abstract}
We present a dedicated analysis of  
the influence of excited states on the calculation of 
nucleon matrix elements. 
This calculation is performed at a fixed value of the lattice spacing, volume and pion mass
that are typical of contemporary lattice computations.
We focus on the nucleon axial charge, $g_A$, for which we use about 7,500 measurements,
and on the average momentum of 
the unpolarized isovector parton distribution, $\left\langle x \right\rangle_{u-d}$, 
for which we use about 23,000 measurements.
All computations are done employing $N_f=2+1+1$ maximally-twisted-mass Wilson fermions and 
using non-perturbatively calculated renormalization 
factors. 
Excited state effects are shown to be negligible for $g_A$, whereas they lead to an $\mathcal{O}(10\%)$ downward shift for $\left\langle x \right\rangle_{u-d}$.
\end{abstract}


\maketitle


\section{\label{sec:int}Introduction}

Understanding nucleon structure is one of the fundamental goals
of lattice QCD. 
Such an endeavor is becoming more realistic as 
present day calculations are being performed  
closer to the limit of physical 
quark masses, small lattice spacings and large volumes~\cite{Hoelbling:2011kk}. 
Thus a direct comparison of results from lattice calculations 
and experimental measurements is becoming feasible, allowing us to probe 
QCD as the underlying theory of the strong interactions.

In order to establish that lattice QCD can indeed provide results 
that address this challenge, we are focusing, in this work, on a number of 
benchmark observables for which QCD 
is expected to produce the right results. 
However, even with the 
significant advances of the past few years,
there is presently an 
unexpected discrepancy between lattice calculations
and experimental measurements 
of physical observables such as the nucleon axial 
charge, $g_A$, the average momentum of
the unpolarized isovector parton distribution, 
$\left\langle x \right\rangle_{u-d}$, and the 
charge radius of the nucleon~\cite{Alexandrou:2010cm,Renner:2010ks,Alexandrou:2011db,Collins:2011mk}. 
Clearly, these discrepancies need an explanation.  
Naturally, this demands a careful study of the systematic effects that  
play an important role in lattice calculations, 
i.e.\ lattice artifacts, finite volume effects and 
non-perturbative renormalization.
Indeed, these systematic effects
are currently being addressed by many
lattice QCD collaborations.
In this work, we focus on the investigation for
two key observables, namely 
$g_A$ and  $\left\langle x \right\rangle_{u-d}$.
These quantities are determined on the lattice by computing suitable ratios
of 3-point and 2-point correlation functions that reach a constant \emph{plateau} value
for asymptotically large Euclidean time separations.

There are several studies of $g_A$ and  $\left\langle x \right\rangle_{u-d}$   
at different values of the 
lattice spacing and various volumes~\cite{Alexandrou:2011nr,Aoki:2010xg,Bratt:2010jn,Pleiter:2011gw}.   
For the values of the pion masses considered, 
these results reveal that taking into account lattice spacing and 
finite volume effects is probably not sufficient 
to reconcile the lattice calculations with the 
current experimental value of $g_A$ 
and the phenomenological extractions of $\left\langle x \right\rangle_{u-d}$. 
In addition, there is a recent study indicating that the discrepancy 
persists even for pion masses quite close to the physical value~\cite{Pleiter:2011gw}.
One might still argue that calculations performed at precisely
the physical point might eliminate these discrepancies but that would require
a strong pion mass dependence making such an explanation
increasingly less plausible.

An issue that is currently under scrutiny within the lattice community concerns 
excited state contamination. These are sub-leading contributions
in Euclidean time correlation functions that can cause a systematic effect in
determining the desired nucleon-nucleon matrix element.
In order to understand whether the physical plateau appears at 
larger Euclidean times than used in present day calculations, we 
need to examine the relevant 3-point functions at 
Euclidean times that are as large as possible.
The difficulty associated with probing such sub-leading contributions to these ratios is 
that the signal-to-noise ratio decreases exponentially fast with Euclidean time.
To tackle this problem, we have performed a dedicated high precision calculation  
of the 3-point functions for $g_A$ 
and $\left\langle x \right\rangle_{u-d}$. In particular, we have used 
about $7500$ measurements for $g_A$ and about
$23000$ for $\left\langle x \right\rangle_{u-d}$.
With such large numbers of measurements, we are able to calculate
the correlation functions for $g_A$ and  $\left\langle x \right\rangle_{u-d}$
for large Euclidean times to sufficient accuracy. This allows us
to study carefully possible excited state contributions, which is the goal
of this work.

This paper is organized as follows.  In section~\ref{sec:NME},
we describe briefly how the matrix elements for the observables of
interest in this work are calculated in 
Euclidean field theory and how the excited state contributions enter the calculation. 
In section~\ref{sec:techniques}, we give the details of the calculation and 
explain the \emph{open sink method} used in this dedicated analysis. 
In section~\ref{sec:results} we present our results
 and in section~\ref{sec:summary} we summarize and conclude.


\section{\label{sec:NME}Nucleon Matrix Elements and Excited State Contributions}

In order to make the paper self-contained, we   
introduce the basic definitions of the quantities studied here.
In the following discussion, we assume an infinite volume, 
while in all our practical calculations, periodic or anti-periodic boundary 
conditions are taken as needed.  The effects of a finite space-time
lattice are felt in the standard finite-size effects of the matrix elements
and the so-called thermal contributions that distort the Euclidean
time dependence of correlation functions.  These systematic effects are
both addressed by the finite-size studies discussed previously and 
are henceforth ignored. 

The zero momentum nucleon 2-point correlation function on the lattice is defined as 
\begin{align*}
 C_2\left( t \right) &= \Gamma_{\alpha \alpha^\prime} \sum\limits_{\vec{x}} 
  \left\langle J_{N, \alpha^\prime}(\vec{x}, t) \, \bar{J}_{{N}, \alpha} (0) \right\rangle
\end{align*}
where $\bar{J}_N$ is a nucleon interpolating field 
that creates a state on the lattice with the same quantum numbers as the nucleon, 
$\Gamma$ is a matrix acting in Dirac space, and  
the sum over the Dirac indices $\alpha$ is implicitly understood.
The Euclidean time $t$ denotes the separation between the creation 
time $t_\mathrm{source}$ 
and annihilation time $t_\mathrm{sink}$ of the nucleon and   
is often referred to as the \emph{source-sink separation},
$t=t_\mathrm{sink}-t_\mathrm{source}$. 
We use translational invariance to set $t_\mathrm{source}=0$ resulting in $t=t_\mathrm{sink}$. 
The nucleon interpolating field is
\begin{align*}
 J_{N,\alpha} (x) &= \varepsilon^{abc} u_\alpha^a (x) \left( \left( d^b (x) \right)^T C \gamma_5 \:  u^c (x)  \right),
\end{align*}
where $C=i \gamma_0 \gamma_2$ is the charge conjugation operator. 
The transfer matrix formalism allows us to relate lattice correlation
functions to matrix elements of operators.  Application of the standard
methods gives the spectral representation of the 2-point function
in terms of the eigenstates of the transfer matrix or equivalently
the Hamiltonian $\mathcal{H}$.  The resulting expression is
\begin{align*}
C_2\left( t \right)
&= \sum\limits_{k} J_{N}^{(k)} \bar{J}_{{N}}^{(k)} \text{e}^{-m_k t}\;,
\end{align*}
which in the limit $t \rightarrow \infty$, will be dominated by the nucleon ground state,
\begin{equation}
 \lim\limits_{t \rightarrow \infty} C_2 \left( t \right) 
\rightarrow J_{N}^{(0)}  \bar{J}_{N}^{(0)} \text{e}^{-m_N t} \; .
\label{2-point}
\end{equation}
In these expressions, $m_k$ labels the masses in the nucleon channel 
and $m_N$ denotes specifically the nucleon mass.
Additionally, we have introduced the symbols 
$J_{N}^{(k)}$ and $\bar{J}_{N}^{(k)}$ for the overlap of the interpolating 
fields with the $k^\text{th}$ eigenstate of $\mathcal{H}$.
Strictly speaking, the limit $t \rightarrow \infty$ cannot be realized on a 
finite lattice; 
in practice however, it suffices to take $t$ large enough
so that the correction coming from the lowest lying excited state can 
be neglected. 
For the evaluation of nucleon matrix elements that we are interested in, 
we additionally need to calculate 
3-point correlation functions. They are defined as
\begin{equation}
 C_3\left( t, t^\prime \right) 
 = \Gamma^\prime_{\alpha \alpha^\prime} \sum\limits_{\vec{x}, \vec{y}} 
    \left\langle J_{N, \alpha^\prime}(\vec{x}, t) \,
      \mathcal{O}(\vec{y}, t^\prime) \,
    \bar{J}_{N, \alpha} (\vec{0},0) \right\rangle,
\label{3-point}
\end{equation}
where $\mathcal{O}$ is a local field corresponding to the operator $\hat{\mathcal{O}}$ of 
interest and $\Gamma^\prime$ is an appropriately defined
matrix acting in Dirac space. 
We denote by $t'$ the insertion time of the operator under consideration. 
Like the 2-point function, there is a spectral representation that
can be derived from the transfer matrix formalism and reads
\begin{align*}
 C_3\left( t, t^\prime \right) 
 &= \sum\limits_{j,k}
    J_{N}^{(j)} \bar{J}_{N}^{(k)} \text{e}^{-m_j (t-t^\prime)} \text{e}^{-m_k t^\prime} 
    \left\langle j \right| \hat{\mathcal{O}} \left| k \right\rangle\; . 
\end{align*}
The asymptotic limit of large Euclidean time again isolates the nucleon contribution
as follows
\begin{align*}
 \lim\limits_{(t-t^\prime), t^\prime \rightarrow \infty} C_3\left( t, t^\prime \right) 
  &\rightarrow J_{N}^{(0)}  \bar{J}_{N}^{(0)} \text{e}^{-m_N t} 
     \left\langle 0 \right| \hat{\mathcal{O}} \left| 0 \right\rangle\; .
\end{align*}
The desired nucleon matrix element 
$\left\langle 0 \right| \hat{\mathcal{O}} \left| 0 \right\rangle$ 
is then obtained from the asymptotic Euclidean time limit
of the ratio of the 3-point and the 2-point function
\begin{align}
  \label{eqn:asymptotic_limit}
  \left\langle 0 \right| \mathcal{O} \left| 0 \right\rangle 
  &= 
  \lim\limits_{(t-t^\prime),t^\prime \rightarrow \infty}
  \frac{C_3\left( t, t^\prime \right)}{C_2\left( t \right)}\; .
\end{align}
It is the main goal of this paper to investigate how large $t-t^\prime$  and $t^\prime$ should be
so that the contribution of  the lowest lying excited state
 -- understood as a systematic error -- becomes negligible within
the desired precision.

Let us begin by discussing in what ways the excited states of the nucleon contribute 
to the nucleon matrix element calculated from the ratio of the 3-point to the 2-point function.
The expressions given in Eqs.~(\ref{2-point}) and (\ref{eqn:asymptotic_limit}) illustrate
how calculations of Euclidean time correlation functions
can be used to determine matrix elements in the limit of large Euclidean time separations.  
However, in practice the finite time extension of the lattice
prevents us from taking the asymptotic limits
and therefore one has to carefully examine the sub-leading contributions usually ignored.
In the following we assume that at the values of $t$ we use, excited state contributions
in the 2-point function can be ignored. This is plausible since excited state contributions 
to the 2-point functions are generically more suppressed 
than those contributions to the 3-point function.
The reasoning for this is that the fields in the 2-point function are always separated by
a distance $t$ and those in the 3-point function are separated by $t-t^\prime$ or
$t^\prime$, both of which are smaller than $t$ in practice.
In particular, in the 3-point function we have a double limit requiring both
 $t-t^\prime$ {\it and} $t^\prime$ to be asymptotic. 
Let us now consider the leading contributions to the ratio of the 3-point and the 2-point functions
originating from taking into account the contribution of the first excited state:
\begin{align*}
 \frac{C_3\left( t, t^\prime \right)}{C_2 \left( t \right)}
   &=
 \quad \left\langle 0 \right| \mathcal{O} \left| 0 \right\rangle \\
&+     \left\langle 0 \right| \mathcal{O} \left| 1 \right\rangle 
       \frac{\bar{J}_{{N}}^{(1)}}{\bar{J}_{{N}}^{(0)}} 
       \: \exp\left(-\Delta M t^\prime\right) \\
&+     \left\langle 1 \right| \mathcal{O}\left| 0 \right\rangle 
       \frac{J_N^{(1)}}{J_N^{(0)}} 
       \: \exp\left[-\Delta M (t - t^\prime)\right]  \\
&+ \mathcal{O}\left[\exp\left(-\Delta M t\right)\right] \; ,
\end{align*}
where $\Delta M$ is the mass gap between the nucleon ground state and the 
first excited state. 
As can be seen, there are two additional time dependent
contributions to leading order.
\section{\label{sec:techniques}Lattice techniques and details}

\subsection{Lattice fermion action}

For this work, we employ maximally-twisted-mass Wilson fermions~\cite{Frezzotti:2003ni}. 
We use the gauge field configurations generated by   
the European Twisted Mass Collaboration (ETMC)
with $N_f=2+1+1$ flavors, thus fully accounting for the first two 
quark generations. 
We refer to \Ref{\cite{Baron:2010bv}} for the details 
of our lattice formulation. 
Since we are aiming at a precise result,
we concentrate on only one ensemble
with a pion mass of $m_\pi\approx 380~\mathrm{MeV}$ and a
lattice spacing of $a \approx 0.078~\text{fm}$.
This pion mass is chosen small enough to be relatively 
close to the physical pion mass but still large enough to
ensure that finite size effects are suppressed. In addition, the propagators 
can be calculated with moderate computational cost which allows us 
to analyze a large statistical ensemble in order to obtain an accurate result.
We emphasize that 
maximally-twisted-mass fermions realize an {\em automatic} $O(a)$-improvement 
for which no additional
operator specific improvements are needed. 
Therefore, 
at the lattice spacing employed here, 
one expects that discretization 
effects are also suppressed. 
This is confirmed by direct calculations  of these matrix elements 
at three different lattice spacings smaller than $0.1~\mathrm{fm}$ 
in Refs.~{\cite{Alexandrou:2011nr,Alexandrou:2010hf,Alexandrou:2010cm}}.

Although in this analysis we use twisted-mass fermions, 
the most important aspects of excited state contributions
are expected to be universal and independent of the particular
lattice discretization used. 
Thus the conclusions obtained here are of direct relevance 
to the calculations of other groups including those using different  
lattice actions.

\subsection{Fixed Sink Method}

\begin{figure}
\centering
\includegraphics[width=0.48\textwidth]{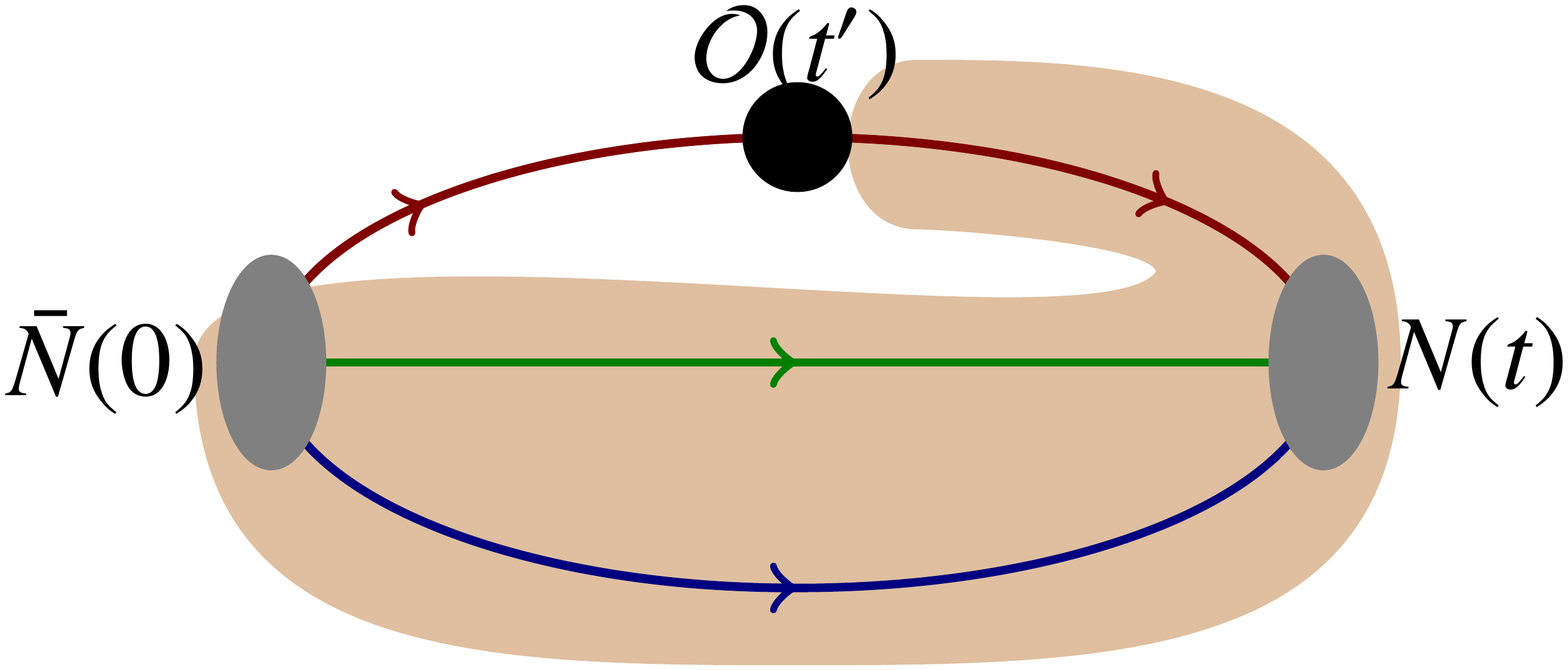}
\includegraphics[width=0.48\textwidth]{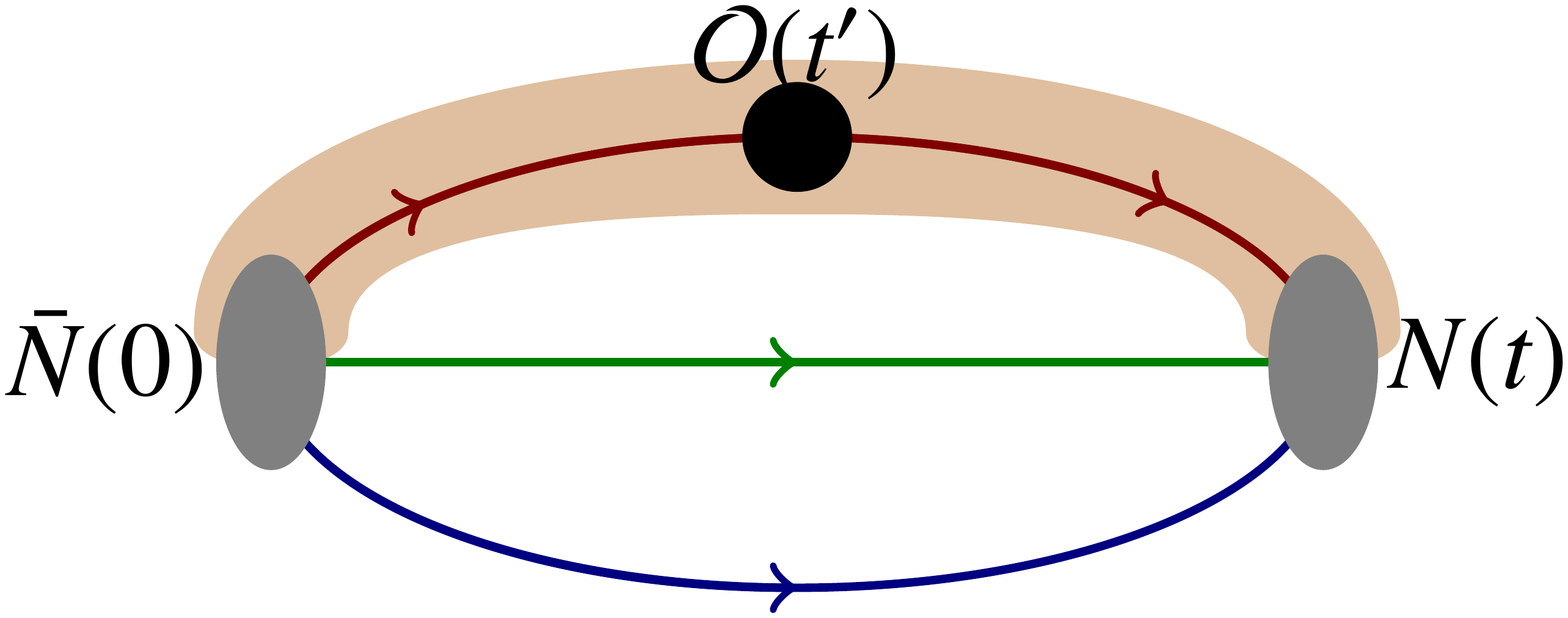}
\caption{\label{fig:diagStandardAndOpenSink}
Diagrammatic illustration of the sequential method through the sink (left) and the open sink method (right).}
\end{figure}

An efficient computation of the connected piece of the 3-point function is possible by means of 
the so-called sequential source method. 
This technique requires two sets of propagators.
The first are the forward quark propagators of the light flavors that are also used
to compute the 2-point function and are independent of the operator 
and hadronic state as long as the same source is used.
Those propagators are then used to build a sequential source for a second
\emph{generalized} propagator, again for each of the light flavors, that is
specific for the hadron state we are interested in.
The desired matrix element is obtained by 
contracting the corresponding operator with the \emph{free ends}
of these two propagators as illustrated in 
\Fig{\ref{fig:diagStandardAndOpenSink}}.
This method has the advantage that we can use the same set of propagators 
for any choice of the operator insertion and hence it is the method
of choice for the calculation of generalized form factors of a particular
hadron.
A disadvantage of this method is that the source-sink separation time $t$ 
in \Eq{\ref{eqn:asymptotic_limit}} must be fixed before the sequential 
propagator can be computed. Obviously, changing $t$ would require 
another set of propagator computations.
Thus, a prudent choice for the value of $t$ is mandatory. 
This is because, on the one hand, according to \Eq{\ref{eqn:asymptotic_limit}}, a large source-sink separation 
is desirable for the suppression of excited state contributions. 
On the other hand, the signal-to-noise ratio drops exponentially fast with the source-sink separation. 
A reasonable choice is therefore a source-sink separation at which the contribution from 
excited states becomes negligible compared to the statistical error. 
However, this can only be determined a posteriori and only after repeating the
entire calculation for several values of $t$. 
Thus in practice, having chosen a reasonable value for $t$, one looks at the time dependence of the right hand 
side of \Eq{\ref{eqn:asymptotic_limit}} as a function of the operator
insertion time $t^\prime$.
If a plateau, as a function of $t'$, is observed, 
then it is assumed that the excited states have been sufficiently suppressed and
the plateau value is identified as the matrix element of interest.
However, there still remains the possibility that the asymptotic plateau value
has not been reached. 
Therefore in this paper, we carry out a thorough investigation of the excited
state contributions using a more appropriate approach as described in the
next section.

\subsection{\label{sec:osm}Open Sink Method}

In the fixed sink method the summation over $\vec{x}$ in the 3-point function 
of Eq.~(\ref{3-point}) is done by the sequential inversion. One can
instead perform the summation over $\vec{y}$ through a sequential inversion.
Therefore we need to fix the particular operator that we are interested in 
and also the time slice  $t^\prime$
where it is inserted. The sequential source is then constructed at the operator
rather than the sink and hence, the sequential propagator 
is now operator-dependent but state and $t$ independent. 
For the example of an  operator that involves only one lattice site, 
the sequential propagator is given by
\begin{align*}
  S^{a^\prime a}_{\alpha^\prime \alpha}(\vec{x_f},t;\vec{x_i},0;\mathcal{O}) &= 
  \sum\limits_{\vec{y}} S^{a^\prime c}_{\alpha^\prime \gamma}(\vec{x_f},t; \vec{y}, t')\: 
  \mathcal{O}^{c b}_{\gamma \beta}(\vec{y}, t')\: S^{b a}_{\beta \alpha}(\vec{y}, t';\vec{x_i},0)
\end{align*}
where the Roman superscripts are color indices and the Greek subscripts are Dirac indices. 
The generalization of this expression for 
operators that involve more than one lattice site and gauge links, 
such as derivative operators, is straightforward.
The sequential propagator is obtained for all
source sink separations $t$, thereby allowing us to study the effects of
excited states.
The open sink or fixed current method is illustrated in \Fig{\ref{fig:diagStandardAndOpenSink}}.
Clearly, this method is not practical when a large number of 
matrix elements of different operators must be computed since for each new 
operator extra inversions must be performed. 
However, for our dedicated study of excited state effects in $g_A$ and $\langle x \rangle_{u-d}$ it 
is clearly very useful. 

\subsection{\label{sec:obs}Observables}

As stated above,
in this work, we concentrate on two relatively simple but nonetheless 
phenomenologically very relevant quantities. 
The first is the nucleon axial charge, $g_A$, which plays   
an important role in the beta decay of the 
neutron and appears as a low energy constant 
in effective chiral Lagrangians.
It has been  precisely measured and it is straightforward to calculate
in lattice QCD using the techniques described in the previous sections.
However, the values obtained from various lattice QCD calculations 
are typically $5\%$ to $10\%$ lower while having themselves a 
statistical accuracy of the order of $1\%$,
see \Fig{\ref{fig:rel_deviation_Nf2and2p1p1}} for the example of our own calculations.

The second observable is the lowest non-trivial moment of the unpolarized 
parton distribution function in isovector flavor combination,
$\left\langle x \right\rangle_{u-d}$. 
This quantity is determined phenomenologically from 
a global analysis of deep inelastic scattering data,
 and the discrepancy between the phenomenological 
and lattice values is even larger, roughly $50\%$ to $60\%$, 
see \Fig{\ref{fig:rel_deviation_Nf2and2p1p1}}.
For the precise definitions of the corresponding 3-point functions,
we refer to \Refs{\cite{Alexandrou:2011db,Alexandrou:2011nr}}.
Though this does not affect excited state contamination, it is important
to note that we use 
a non-perturbative renormalization of our bare matrix elements. 
The corresponding renormalization factors are calculated in the 
$\text{RI}^\prime\text{MOM}$ scheme and are matched to the $\overline{\text{MS}}$ 
scheme at a scale of $(2~\text{GeV})^2$.
For more details we refer to \Refs{\cite{Alexandrou:2010me,Dinter:2011jt}}. 
The values of the renormalization constants used in this work are $Z_A = 0.774$ for the 
renormalization of the bare $g_A$ and $Z_{\langle x\rangle}= 0.998$ for the renormalization 
of $\left\langle x \right\rangle_{u-d}$. 

\section{\label{sec:results}Results}

\begin{figure}[ht]
\includegraphics[width=0.48\textwidth]{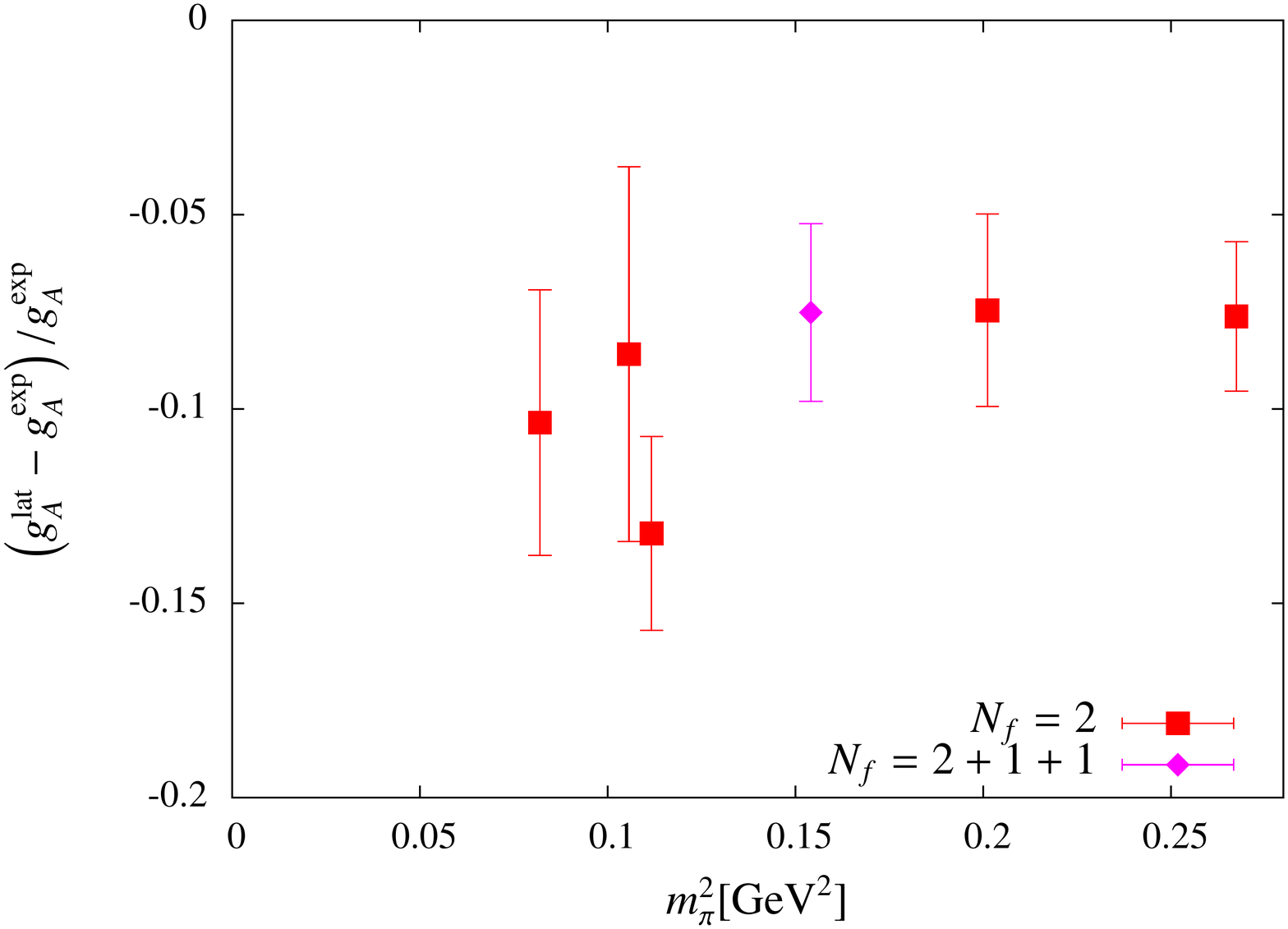}
\includegraphics[width=0.48\textwidth]{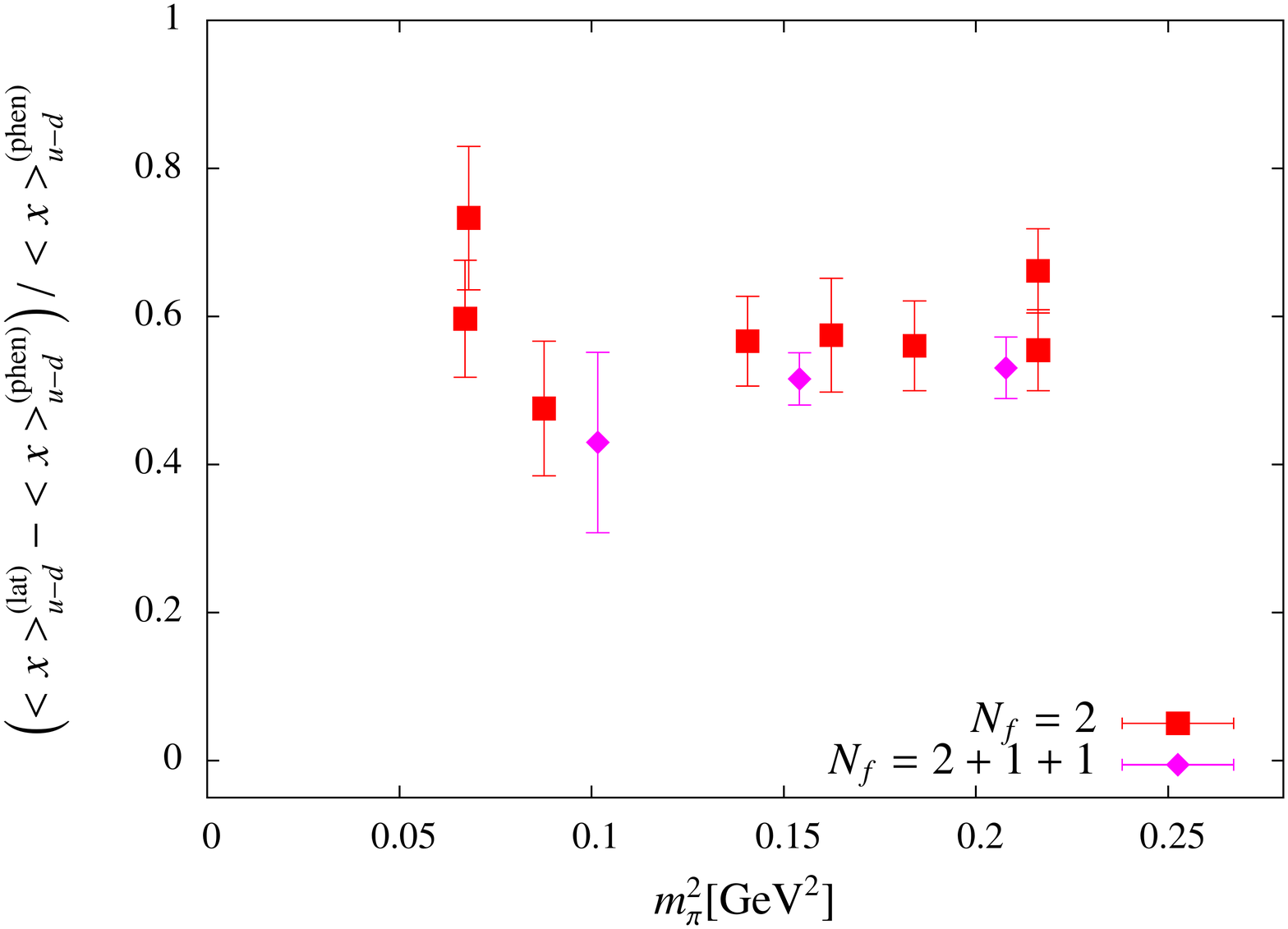}
\caption{\label{fig:rel_deviation_Nf2and2p1p1}
In the left panel, we show the relative deviation of 
ETMC lattice results for $g_A$ from the experimental value~\cite{Nakamura:2010zzi}. 
In the right panel, we show the relative deviation of 
ETMC lattice results for $\left\langle x \right\rangle_{u-d}$
from a result obtained from a phenomenological analysis~\cite{Alekhin:2009ni}.
The lattice values for $N_f=2$ at the various pion masses
 are from Refs.~\cite{Alexandrou:2010hf,Alexandrou:2011nr}. 
The filled (magenta) diamonds show the results using the $N_f=2+1+1$ ensembles.} 
\end{figure}

In order to have a reference value, we  first perform a calculation  
of the nucleon axial charge $g_A$ using the 
sequential fixed sink method with a fixed source-sink separation 
of \nopagebreak{$t=12a\approx 0.94~\text{fm}$}. 
Gauge invariant Gaussian smearing of the quark fields, including APE-smeared gauge links, 
is used in order to improve  
the overlap with the nucleon ground state. 
In \Fig{\ref{fig:rel_deviation_Nf2and2p1p1}} we compare the value obtained for 
the $N_f=2+1+1$ ensemble to 
our previous $N_f=2$ results at various pion masses.
As can be seen, the value we find for $g_A$ is in  good agreement with
the $N_f=2$ results obtained at nearby pion masses.
 
\begin{figure}[ht]
\includegraphics[width=0.9\textwidth]{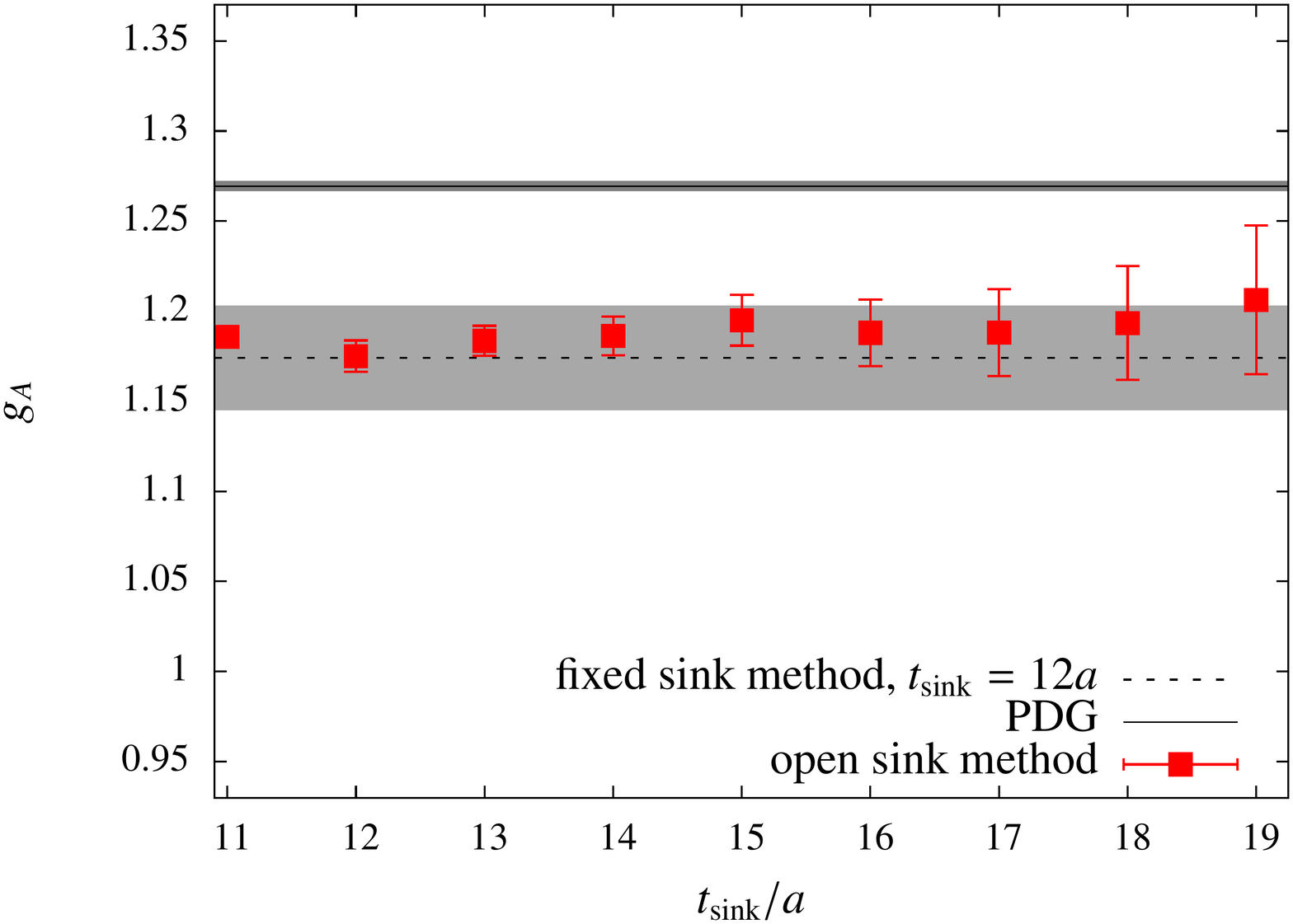}
\caption{\label{fig:gA_open_sink} Results for $g_A$ for a range of source-sink separations 
obtained from the open sink analysis on one $N_f=2+1+1$ ensemble.
The light grey band indicates the result obtained from the fixed sink method  
using a source-sink separation of $12a$ and the dark grey band shows the experimental value.}
\end{figure}

We then perform an analysis on the same $N_f=2+1+1$ ensemble using the open sink method. 
The time slice of the operator insertion was fixed to $t'=9a$. This was chosen
to safely suppress excited state contributions from the source, as can be
verified from the 2-point function.
The result of the analysis using the open sink method is shown in 
\Fig{\ref{fig:gA_open_sink}}. 
Clearly, 
the value of $g_A$ does not show any statistically significant dependence on 
the source-sink separation $t$. Hence,  
the plot demonstrates that there is no 
contribution from excited states detectable within the statistical accuracy 
of 2.5\%. 
Note that, although $t'=9a$, 
a plateau for $g_A$ is already reached  at $t=11a$. 
It is worth mentioning that, in order to reach a comparable 
statistical accuracy as the one obtained when using the fixed sink method 
with $t=12a$ with 500 measurements, we had to perform 
roughly $7500$ measurements when we take for example $t=18a$.
This was achieved by choosing randomly located source points
with typically $2$ sources per configuration.

\begin{figure}[ht]
 \includegraphics[width=0.9\textwidth]{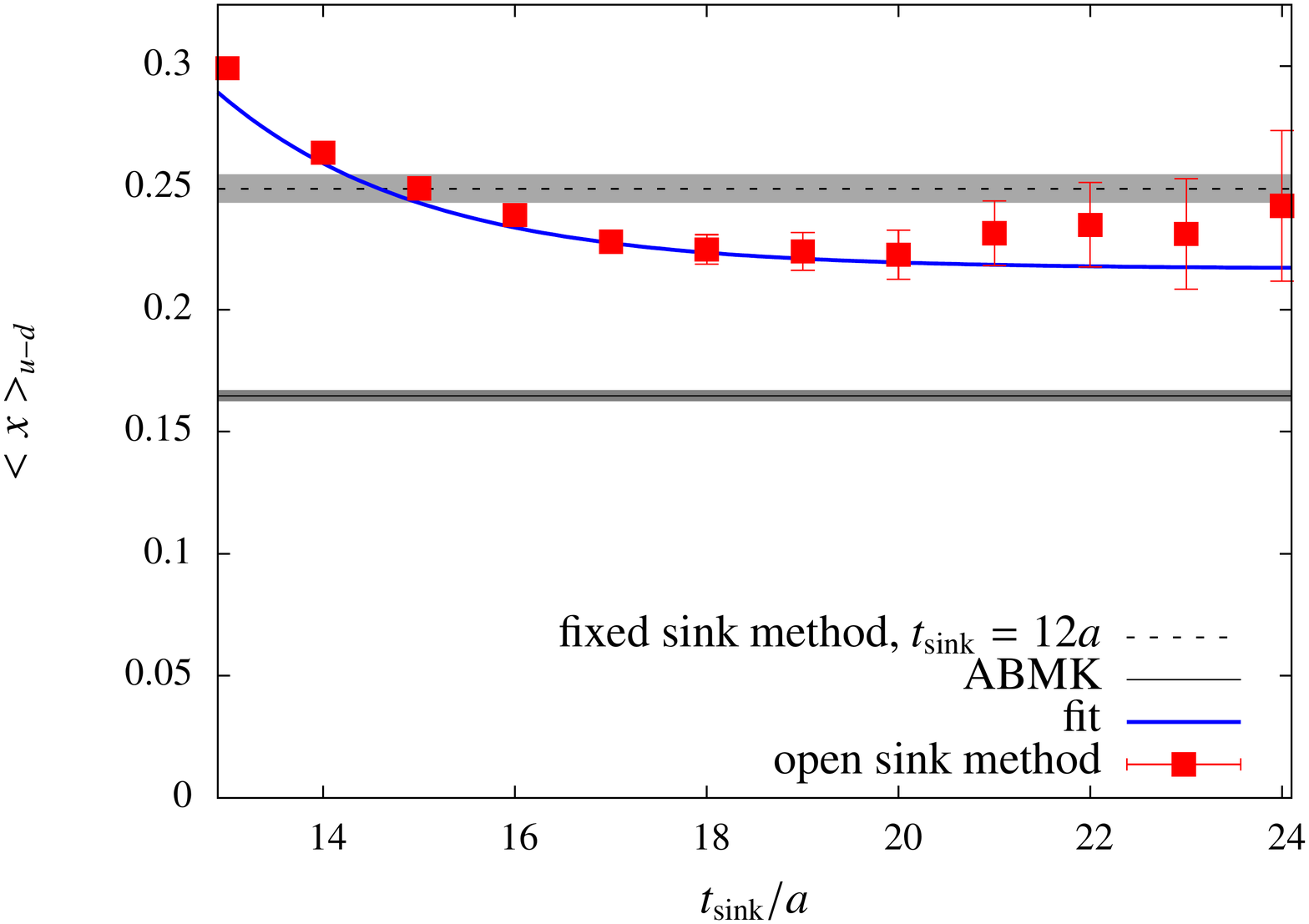}
 \caption{\label{fig:averageX_open_sink}
Results for 
$\left\langle x \right\rangle_{u-d}$ for a range of 
source-sink separations obtained by means of the open sink method. 
The operator insertion was at a 
temporal separation from the source of $t'=11a$. 
The value (including errors) obtained from the fixed sink method
using a source-sink separation of $12a$ is indicated by the light grey band.
The phenomenologically extracted value is shown with the dark grey band. 
The blue solid line corresponds to a fit described in the text.}
\end{figure}

As a second benchmark quantity, we have examined
$\left\langle x \right\rangle_{u-d}$.
As for the case of $g_A$, we first determine
the value of this observable using the fixed sink method 
increasing the statistics of the calculation 
presented in~\cite{Dinter:2011jt}. 
For the open sink method, we have chosen the operator  
insertion time to be $t^\prime=11a$. 
We expect that for this choice excited state effects 
from the source are sufficiently suppressed for this operator.
We perform in total about $23,000$ measurements for 
$\left\langle x \right\rangle_{u-d}$ using randomly distributed source 
points with $5$ sources per configuration.
With such statistics and at a source-sink separation of $t=18a$, 
we could equal the precision 
of the fixed sink method that was done 
with a source-sink separation of $t=12a$ using $1300$ measurements.
In \Fig{\ref{fig:averageX_open_sink}} we plot $\left\langle x \right\rangle_{u-d}$ 
as a function of the source-sink separation $t$.
We also indicate the value obtained from the fixed sink method 
analysis as well as the experimental result extracted from a recent 
global analysis~\cite{Alekhin:2009ni}.

As can be seen, for this observable there is a shift of the value 
of $\left\langle x \right\rangle_{u-d}$ 
and a plateau is reached at larger values of the source-sink separation than what
we have used in the fixed sink method.
Despite the fact that the results for larger values of $t$ decrease
they clearly do not reach the phenomenological value. 
In order to estimate any residual dependence on $t$,
we determined the value of $\left\langle x \right\rangle_{u-d}$
for an infinite source-sink separation 
by fitting the expected exponential behavior,
\begin{align*}
 \label{eq:exp_fit}
\left\langle x \right\rangle_{u-d} + A \exp \left[ -\Delta M\left(t - t'\right) \right] \;,
\end{align*}
to the lattice results with a fixed $t'=11a$. 
The result of this fit is 
$\left\langle x \right\rangle_{u-d} = 0.22(1)$, 
which is $12\%$ lower than the result of $\left\langle x \right\rangle_{u-d} = 0.250(6)$, 
obtained using $t=12a$ in the fixed sink method. 
The error of the fit is estimated by varying the fit range and 
by comparing the use of a fixed parameter $\Delta M$ versus fitting $\Delta M$ directly.


\section{\label{sec:summary}Summary and Conclusions}

In this letter 
we have performed precision calculations of 
$g_A$ and $\left\langle x \right\rangle_{u-d}$ for a single ensemble 
of gauge field configurations with $N_f=2+1+1$ dynamical fermions 
employing a non-perturbative renormalization. 
We have investigated the behavior of these benchmark quantities 
as a function of the source-sink separation in order to 
assess the influence of excited states on the current lattice results for
$g_A$ and $\left\langle x \right\rangle_{u-d}$.  This is particularly
important given that excited states may play a role in
explaining the presently observed discrepancy between lattice computations and 
phenomenological evaluations of several important nucleon observables.

We find that for the here considered pion mass of about
$380~\text{MeV}$ and lattice spacing of $a\approx0.078~\mathrm{fm}$, 
the contamination of excited states 
is negligible for $g_A$, but for 
$\left\langle x \right\rangle_{u-d}$, the effect is of 
the order of $10\%$ compared to our previous calculations,
where the source-sink separation has been set to about $1$~fm. 
This is an effect larger than 
the finite volume and lattice spacing effects we observe
at this value of the pion mass, volume and lattice spacing.
Moreover, this demonstrates that contributions from excited states are
operator dependent and should be investigated separately for each operator. 

One way to better control excited state effects
is to use a variational method
such as the generalized eigenvalue method \cite{Luscher:1990ck,Blossier:2009kd}.
Recently, a new approach to deal with 
excited state contamination of hadronic matrix elements has been developed 
and applied for the $B^\ast B\pi$ coupling in ref.~\cite{Bulava:2011yz}.
Whether the generalized eigenvalue method can improve the
calculation of matrix elements of the nucleon
needs still to be tested, though.

However, if the $10\%$ shift for $\left\langle x \right\rangle_{u-d}$ 
as we found here persists at smaller pion masses,
excited state effects can not be the single dominating systematic effect
responsible for the tension between lattice and phenomenology. 
Of course, we cannot exclude that at smaller values of the pion mass
excited state effects might become significantly larger.
Therefore, in order to clarify the deviation between lattice calculations 
and experimental determinations of nucleon matrix elements, 
a very careful and accurate analysis of systematic errors will be needed, 
taking into account the possible contamination of excited states as
observed in this work. 


\section*{Acknowledgments}

\noindent
This work was performed using HPC resources provided by the
 JSC Forschungszentrum J\"ulich on the JuGene supercomputer.
It is supported in part by  the DFG
Sonder\-for\-schungs\-be\-reich/ Trans\-regio SFB/TR9,
by GENCI (IDRIS-CINES) Grant 2010-052271
and by funding received from the
Cyprus Research Promotion Foundation under contracts EPYAN/0506/08,
KY-$\Gamma$/0907/11/ and 
TECH\-NO\-LO\-GY/$\Theta$E$\Pi$I$\Sigma$/ 0308(BE)/17.
It is additionally coauthored in part by Jefferson
Science Associates, LLC under U.S. DOE Contract No. DE-AC05-06OR23177.
C.A.\ acknowledges  partial support by the Research Executive Agency (REA) 
of the European Union under Grant Agreement 
number PITN-GA-2009-238353 (ITN STRONGnet).


\bibliography{paper}

\begin{thebibliography}{10}

\bibitem{Hoelbling:2011kk}
C.~Hoelbling.
\newblock {Light hadron spectroscopy and pseudoscalar decay constants}.
\newblock {\em PoS}, LATTICE2010:011, 2010.

\bibitem{Alexandrou:2010cm}
C.~Alexandrou.
\newblock {Hadron Structure and Form Factors}.
\newblock {\em PoS}, LATTICE2010:001, 2010.

\bibitem{Renner:2010ks}
D.~Renner.
\newblock {Status and prospects for the calculation of hadron structure from
  lattice QCD}.
\newblock {\em PoS}, LAT2009:018, 2009.

\bibitem{Alexandrou:2011db}
C.~Alexandrou, M.~Brinet, J.~Carbonell, M.~Constantinou, P.A. Harraud, et~al.
\newblock {Nucleon electromagnetic form factors in twisted mass lattice QCD}.
\newblock {\em Phys.Rev.}, D83:094502, 2011.

\bibitem{Collins:2011mk}
S.~Collins, M.~Gockeler, Ph. Hagler, R.~Horsley, Y.~Nakamura, et~al.
\newblock {Dirac and Pauli form factors from lattice QCD}.
\newblock 2011.

\bibitem{Alexandrou:2011nr}
C.~Alexandrou, J.~Carbonell, M.~Constantinou, P.A. Harraud, P.~Guichon, et~al.
\newblock {Moments of nucleon generalized parton distributions from lattice
  QCD}.
\newblock {\em Phys.Rev.}, D83:114513, 2011.

\bibitem{Aoki:2010xg}
Y.~Aoki, T.~Blum, H.-W. Lin, S.~Ohta, S.~Sasaki, et~al.
\newblock {Nucleon isovector structure functions in (2+1)-flavor QCD with
  domain wall fermions}.
\newblock {\em Phys.Rev.}, D82:014501, 2010.

\bibitem{Bratt:2010jn}
J.D. Bratt et~al.
\newblock {Nucleon structure from mixed action calculations using 2+1 flavors
  of asqtad sea and domain wall valence fermions}.
\newblock {\em Phys.Rev.}, D82:094502, 2010.

\bibitem{Pleiter:2011gw}
D.~Pleiter et~al.
\newblock {Nucleon form factors and structure functions from N(f)=2 Clover
  fermions}.
\newblock {\em PoS}, LATTICE2010:153, 2010.

\bibitem{Frezzotti:2003ni}
R.~Frezzotti and G.C. Rossi.
\newblock {Chirally improving Wilson fermions. 1. O(a) improvement}.
\newblock {\em JHEP}, 0408:007, 2004.

\bibitem{Baron:2010bv}
R.~Baron, Ph. Boucaud, J.~Carbonell, A.~Deuzeman, V.~Drach, et~al.
\newblock {Light hadrons from lattice QCD with light (u,d), strange and charm
  dynamical quarks}.
\newblock {\em JHEP}, 1006:111, 2010.

\bibitem{Alexandrou:2010hf}
C.~Alexandrou et~al.
\newblock {Axial Nucleon form factors from lattice QCD}.
\newblock {\em Phys.Rev.}, D83:045010, 2011.

\bibitem{Alexandrou:2010me}
C.~Alexandrou, M.~Constantinou, T.~Korzec, H.~Panagopoulos, and F.~Stylianou.
\newblock {Renormalization constants for 2-twist operators in twisted mass
  QCD}.
\newblock {\em Phys.Rev.}, D83:014503, 2011.

\bibitem{Dinter:2011jt}
S.~Dinter, C.~Alexandrou, M.~Constantinou, V.~Drach, K.~Jansen, and D.~Renner.
\newblock {Nucleon matrix elements with $N_f=2+1+1$ maximally twisted
  fermions}.
\newblock {\em PoS}, LATTICE2010:135, 2010.

\bibitem{Nakamura:2010zzi}
K.~Nakamura et~al.
\newblock {Review of particle physics}.
\newblock {\em J.Phys.G}, G37:075021, 2010.

\bibitem{Alekhin:2009ni}
S.~Alekhin, J.~Blumlein, S.~Klein, and S.~Moch.
\newblock {The 3, 4, and 5-flavor NNLO Parton from Deep-Inelastic-Scattering
  Data and at Hadron Colliders}.
\newblock {\em Phys.Rev.}, D81:014032, 2010.

\bibitem{Luscher:1990ck}
Martin Luscher and Ulli Wolff.
\newblock {How to calculate the elastic scattering matrix in two-dimensional
  quantum field theories by numerical simulation}.
\newblock {\em Nucl.Phys.}, B339:222--252, 1990.

\bibitem{Blossier:2009kd}
Benoit Blossier, Michele Della~Morte, Georg von Hippel, Tereza Mendes, and
  Rainer Sommer.
\newblock {On the generalized eigenvalue method for energies and matrix
  elements in lattice field theory}.
\newblock {\em JHEP}, 0904:094, 2009.

\bibitem{Bulava:2011yz}
J.~Bulava, M.~Donnellan, and R.~Sommer.
\newblock {On the computation of hadron-to-hadron transition matrix elements in
  lattice QCD}.
\newblock 2011.

\end{thebibliography}

\end{document}